\documentstyle[aps,twocolumn,epsfig]{revtex}

\begin{document}
\newcommand{\be}{\begin{equation}}
\newcommand{\ee}{\end{equation}}
\newcommand{\bea}{\begin{eqnarray}}
\newcommand{\eea}{\end{eqnarray}}
\newcommand{\bdm}{\begin{displaymath}}
\newcommand{\edm}{\end{displaymath}}
\newcommand{\moy}[1]{\left < #1 \right >}
\newcommand{\pr}[1]{\left ( #1 \right )}
\newcommand{\eqref}[1]{(\ref{#1})}
\bibliographystyle{unsrt}
\title{On the Minority Game : Analytical and Numerical Studies}
\author{Damien Challet and Yi-Cheng Zhang}
\address{Institut de Physique Th\'eorique, Universit\'e de Fribourg, 1700 Fribourg}
\maketitle

\abstract{\it{We investigate further several properties of the {\em minority game} we have recently introduced. We explain the origin of the phase transition and give an analytical expression of $\sigma^2/N$ in the $N\ll2^M$ region.
The ability of the players to learn a given payoff is also analyzed, and we show that the Darwinian evolution process tends to a self-organized state, in particular, the life-time distribution is a power-law with exponent -2. Furthermore, we study the influence of identical players on their gain and on the system's performance. Finally, we show that large brains always take advantage of small brains.}}
\vspace{3ex}

Recently we have studied a simple model of a minority game \cite{CZ} that captures the essential features of the ``bar-problem'' of Arthur (\cite{Arthur} and \cite{Johnson}). $N$ players compete with each other and act by induction and adaptation. They must choose one side between the two at each time step and those who happen to be in the minority side win. They receive a reward when making the right choice, keep in memory the $M$ last sides which were the right ones and use this knowledge to act at the next time step. Each player possesses a finite set of $S$ strategies and uses the one which would have been the most rewarding if it had been used since the beginning. A strategy is a behavior rule that stipulates an action for every information possible (see \cite{CZ} or \cite{YCZ} for more details). Recently, Savit, Manuca and Riolo \cite{Savit} have studied the step payoff case where each player has the same memory $M$ and only two strategies. They have found a phase transition with parameter $\rho=2^M/N$.

Here, we  continue the study of the payoff learning and the Darwinism's process initiated in \cite{CZ} and introduce other interesting sights of the model. We also give a geometrical explanation of the phase transition and find an analytical expression of $\sigma^2/N$ in the $N\ll2^M$ region.

\section{Learning of different payoffs}

\begin{center}
	\begin{figure}
		
		\psfig{file=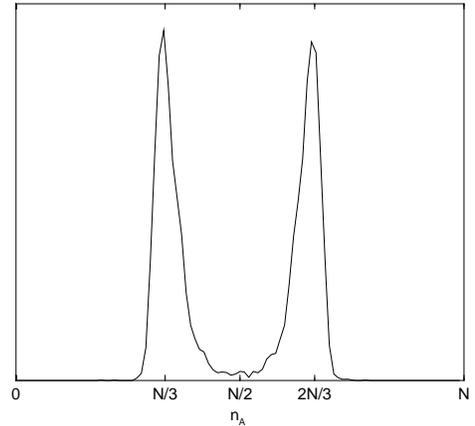,width=8cm}
		\caption{\label{cloche}Histogram of the attendance $n_A$ at side $A$ ($N=101$, $M=6$, $S=15$). The system has maximized the global payoff $w^2(N/2-w)$ by getting the typical number of winners equal to N/3}
	\end{figure}
\end{center}

The simplest payoff consists in giving a point for every good choice, i.e. the individual payoff function is equal to one, the global payoff is equal to $w$, the number of winners. It fits the minority game, but in most other real situations the winners must share limited resources. For example, consider the global payoff $F_{a,b}(w)=aw+b$.  By varying $a$ and $b$, we can reproduce some real situations; for instance, the lottery game corresponds to $a=0$, $b>0$ and the so-called step payoff function is obtained with $a>0$ and $b=0$. In general, a player wishes to maximize his profit, i.e. a priori the individual payoff function $f(w)$.  It is tempting to gain the maximum of $f(w)$ at each time step, but it occurs in average only every $N/\moy{w}$ time steps, thus the system is expected to try to get the average number of winners giving the maximum of the global payoff function $F(w)=f(w)w$, say $w_{max}$. The  $w_{max}=1$ case reveals one limit of the model due its binary nature. Each player would like to win alone, but he has to face the reality : the minimum number of winners is equal to $N/2^S$ in average, whose all strategies give the same answer to this information. If $S=2$ and $w_{max}=N/2$ it is a security, but it is harmful when $N/2^s\geq w_{max}$. If $w_{max}\sim N/2^S$,  $N/2^S$ players sometimes win huge rewards, but it only happens intermittently and always after the same information, because the other players are blinded by the enormous virtual gain; thus there are only two groups of roughly $N/2^S$ people that can hit the jackpot. The morals is the following : one can win a lot, but not intentionally. If the global payoff has a maximum in $w_{max}$ greater enough than $N/2^S$, the system can manage to maximize the global payoff. For instance, we take $f(w)=w(N/2-w)$, that is, $F(w)=w^2(N/2-w)$ has a maximum in $N/3$. If $N=101$, $M=6$ and $S=15$, it is clear that the peaks are in $w=N/3$ (see figure \ref{cloche}). 

Nevertheless there are conditions under which the system can get two peaks.  In general, if $\rho>\rho_c$ the histogram of the attendance at side A can only have one peak centered on $N/2$. If $\rho<\rho_c$ the histogram has at least two peaks that are symmetrical with respect to $N/2$ and whose positions strongly depend on $\rho$, and, if $w_{max}=N/2$, a peak centered in $N/2$. Consequently, if $w_{max}\ll N$, a system can maximize the global gain only if $N\gg 2^M$ and $N\ll 2^S$. 

\section{Darwinism}

\begin{center}
	\begin{figure}
		\psfig{file=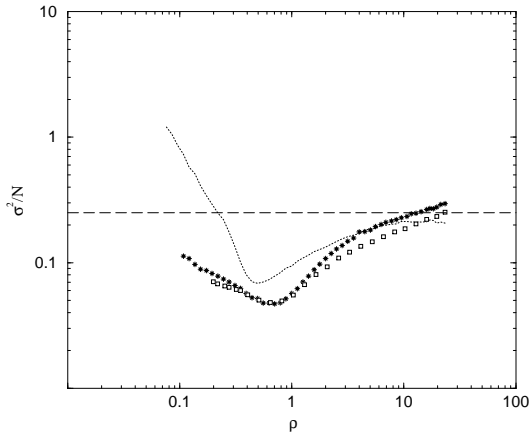,width=8cm}
		\caption{\label{comp} Dependence of $\sigma^2/N$ in $\rho$ with Darwinism for $p=0.01$ (stars) and $p=0.1$ (squares) ($M=5$, $S=2$, $\tau=10$). The dashed line represents the random performance and the dotted line is $\sigma^2/N$ without Darwinism.} 
	\end{figure}
\end{center}

The Darwinism process is the same as in \cite{CZ} : every $\tau$ time steps, the worst player is replaced by a clone of the best, except that one strategy is redrawn with a small probability $p$ in order to allow regeneration, and that the virtual gains of the strategies are reset to zero, like a new born baby. If the new player is a pure clone of the best one, one says that both belong to the same species. A demonstration of the Darwinism's benefits can be seen in figure \ref{comp} : the latter shows a comparison between the variance of the attendance signal with and without Darwinism. The region where $\sigma^2/N<1/4$ is much greater when evolution takes place, in particular the $\rho<\rho_c$ region is less affected by the overcrowding. Note that increasing $p$ lowers $\sigma^2/N$, that is, the mutations are useful. The asymptotical behaviors of $\sigma^2/N$ in the $\rho\rightarrow\infty$ limit depends on $p$, but the Darwinism is harmful in this region. One also sees that the minimum of $\sigma^2/N$ is not as the same $\rho$ than without Darwinism. One can wonder why the coordination is better in the crowded phase although there are multiple clones of a lot of players. Savit et al. \cite{Savit} pointed out the existence of a harmful process of period 2, more precisely, the first time an information is given to the system, $N/2+O(\sqrt{N})$ players choose one given side; the next time the same information occurs, $N/2+O(N)$ players go to the opposite side. When evolution takes part, due to the fact that the new player loses the virtual gains of the strategies he gets, the periodic process fades away because the harmful coordination vanishes.

\begin{center}
	\begin{figure}	
		\psfig{file=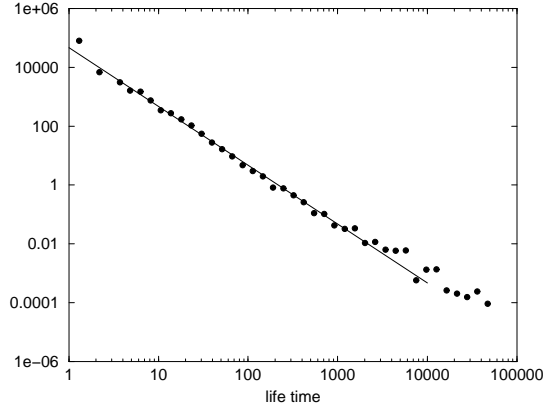,width=8cm}
		\caption{\label{dureevie}Distribution of species life time (N=101, M=8, S=2, NIT=500000, $\tau$=10). The straight line has a -2 exponent. The right part of the distribution does not lie on the same line, because of fluctuations.} 
	\end{figure}
\end{center}

\begin{center}
	\begin{figure}
		\psfig{file=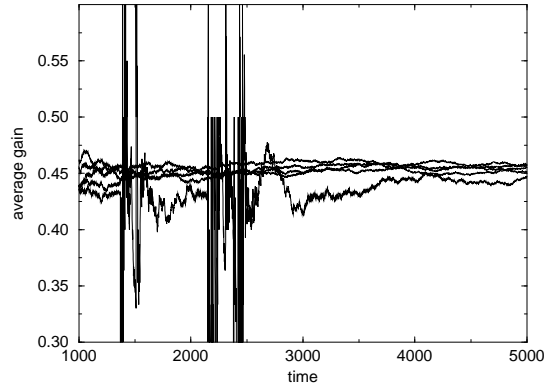,width=8cm}
		\caption{\label{cheveux}Temporal evolution of the average gain of several players ($N=51$, $M=10$, $S=2$, $\tau$=10). Note the consequences of the death of a player at t=1450 and t=2100.} 
	\end{figure}
\end{center}

Since $p<1$, the diversity (the number of different species) is reduced by the evolutionary process, and tends to a value that depends on $p$. Indeed, if $p=1$, the diversity remains equal to $N$; when $p$ decreases, the diversity decreases too, but stays over $N/2$ even if $p=0$. Even more, when one starts with only one species, but with random virtual gains, the system performs first very badly, slowly improves itself, and reaches an optimal diversity always greater than $N/2$. When the diversity is stable, the system is in a stationary state, and one can study several distributions. First, figure \ref{dureevie} shows the distribution of the species life time. It is a power law with exponent $-2.02\pm 0.02$, which does not depend on the parameters  ($\tau$, $p$, ...). Fortunately, it is the same exponent as found in real life evolution \cite{Raup}). Figure \ref{cheveux} helps to understand what happens. The average gain of several players during the game is plotted. One can see players remaining in the game, some other resisting for a while, then disappearing. The player that replaces a dead one is followed. This figure shows that the fluctuations of the average gain is very high when a player is young. Consequently such a player's death or reproduction are more likely than those of an old player; that leads to punctuated equilibrium, explaining the origin of the power law distributions.

\begin{center}
	\begin{figure}		
		\psfig{file=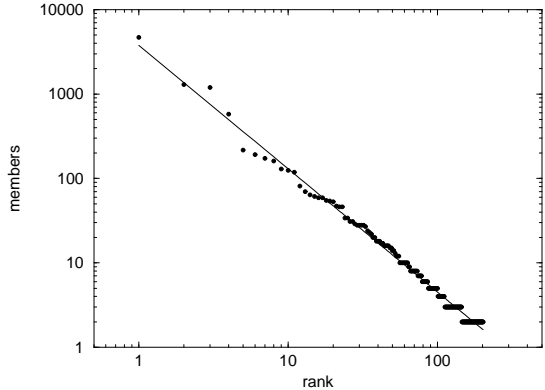,width=8cm}
		\caption{\label{types} Rank of the members in a type of player ($N=20001$, $S=5$, $M=5$, $\tau=10$). It is a power-law, but the exponent depends on the system's parameters.} 
	\end{figure}
\end{center}

After sufficient time, it is interesting to plot the number of members composing each species against their rank (see figure \ref{types}). One find a power law that depends at least on p; indeed, if $p=1$, the best performer will never be completely cloned and one obtains a flat line : every type of player has only one member.

\begin{center}
	\begin{figure}		
		\psfig{file=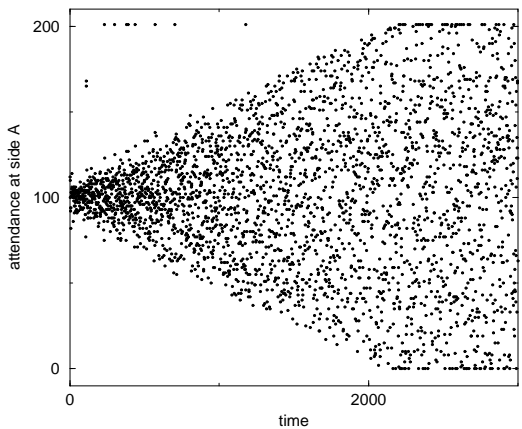,width=8cm}
		\caption{\label{antidarw}Anti Darwinism : a clone of the worst player replaces the best. The attendance at side A is plotted with dots.} 
	\end{figure}
\end{center}

In order to know if the proposed Darwinian process is the good one, we have tried to apply the inverse process : a clone of the worse player replaces the best one. The figure \ref{antidarw} shows that the global performance suffers a lot.

\section{Explanation of the phase transition}

\begin{center}
	\begin{figure}		
		\psfig{file=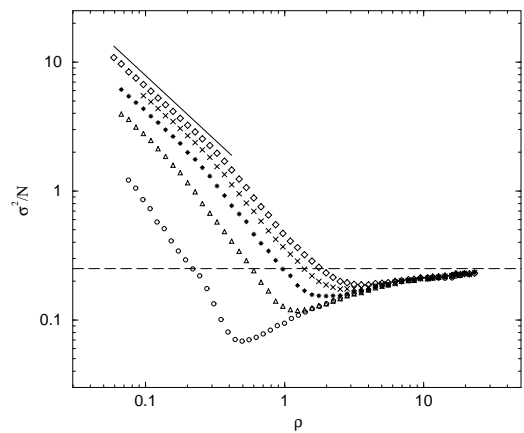,width=8cm}
		\caption{\label{s2Ns}Dependence of $\sigma^2/N$ on $\rho$ for $S=2,3,4,5$ and $6$ (circles, triangles, stars, x and diamonds) ($M=8$). The continuous line is the plot of $1/x$, and the dashed one represents the random case's performance.}
	\end{figure}
\end{center}

Let us now discuss in detail figure \ref{s2Ns} where the phase transition clearly appears. Considering the step payoff function, one has plotted $\sigma^2/N$ against $\rho=2^M/N$ for $S=2,3,4,5$; the dashed line $\sigma^2/N=1/4$ gives the performance that would be attained if every player chose its side randomly. Let us call $\rho_c$ the X-coordinate where $\sigma^2/N$ is minimal. Three regions appear : i) $\rho\ll \rho_c$ : in this region, $\sigma^2/N\sim 1/\rho$, that is, $\sigma^2/N^2\sim 1/2^M$; keeping $M$ constant and increasing $N$ produces only a dilatation of the system, whose origin is discussed below; ii)  $\rho\sim\rho_c$ : the fluctuations are minimal and the system performs better than by coin-tossing; iii) $\rho\gg\rho_c$ : $\sigma^2/N$ tends to 1/4, the random case's performance, because the players are not any more enough to coordinate. 

One sees that increasing the number of strategies reduces the $\rho\sim\rho_c$ region, and that $\sigma^2/N$ depends only on $\rho$ if $\rho$ is large. Furthermore, $\rho_c$ is roughly linear in $S$, as one can see on the following table (values are only approximative) :
\begin{center}
\begin{tabular}{|c|c|}
\hline
$S$&$\rho_c$\\ \hline
2&1/2\\ \hline
3&4/3\\ \hline
4&2 \\ \hline
5&5/2\\ \hline
6&4 \\ \hline
8&6 \\ \hline
\end{tabular}
\end{center}

This figure leads up to two questions. First, since a player completely ignores the details of the other players' strategies, how is it possible that the system performs better than by coin tossing ? Second, why is the parameter $\rho=2^M/N$ the good one, that is, what is the meaning of $\rho$ ? Indeed we know that there are $2^{2^M}$ strategies. One would have expected the phase transition to occur when $N\sim 2^{2^M}$. Why is it not the case ?

First it is clear that the only way the players can interact is the virtual values of their strategies. The coordination occurs because there is some information in these values. The nature of this information is the following. Since a strategy consists in $2^M$ bits, it belongs to an hypercube of dimension $2^M$, $H_M$. For all $s\in H_M$, $s(i)\in\{0,1\}$, $i=1,\dots,2^M$ are the components of $s$. We take the $|.|_1$ distance on this hypercube, also called hamming distance, which counts the number of different bits : let $s$ and $t\in H_M$, then 

$$D_M(s,t)=\sum_{i=1}^{2^M}|s(i)-t(i)|.$$

\begin{center}
	\begin{figure}		
		\psfig{file=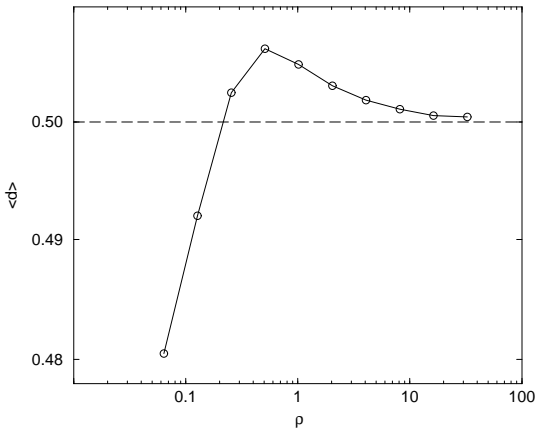,width=8cm}
		\caption{\label{deffZ}Log-log plot of the average actual distance $\moy{d}$ against $\rho$ ($N$=51, $M=2,..,11$, $S=2$).}
	\end{figure}
\end{center}

\begin{center}
	\begin{figure}		
		\psfig{file=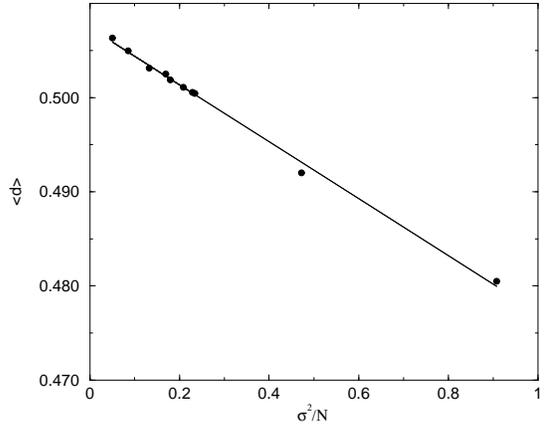,width=8cm}
		\caption{\label{distsigma}Linear relationship between $\sigma^2/N$ and $\moy{d}$. The continuous line shows a linear fit.}
	\end{figure}
\end{center}

For convenience, we introduce the normalized distance $d_M(s,t)=D_M(s,t)/2^M$. Now, given an information, the probability for these two strategies of reacting the same way is given by $1-d_M(s,t)$. Therefore, using the strategy with the highest virtual value is equivalent to choosing the one that has the smallest probability to reacting in the same way as the other players. It is clear now that the virtual value of a strategy is related to its average distance from all strategies used by all other players. Of course each player's dream is to have at each time step at least a strategy that has been in average further than 1/2 from all other players' used strategies. Let us define $\moy{d}$, the average actual distance between the players by computing the average distance between the strategies being used at this moment by the players at each time step, and by averaging this quantity over the whole history of the game. One sees on figure \ref{deffZ} that $\moy{d}$ is actually greater than $1/2$ as soon as $\sigma^2/N<1/4$. Even more, there is a linear relationship between $\moy{d}$ and $\sigma^2/N$, as shown on figure \ref{distsigma}.

There are $2^{2^M}$ different strategies. But take a strategy and change only one bit of it. The distance between these two strategies is $1/2^M$, that is, they are not really different. The question is now : given $s$, a strategy, how many strategies are significantly different from $s$ ? Of course, there is one strategy $\overline{s}$ that is exactly the inverse of $s$ : $\overline{s}(i)=1-s(i)$, $i=1,\dots,2^M$, but one can consider too all strategies which are at a distance of 1/2 from $s$, so to say uncorrelated with $s$. Note that if $s$ and $t$ are uncorrelated, $\overline{s}$ is also uncorrelated with $t$. In appendix A, we show that an ensemble whose all elements are mutually uncorrelated contains at most $2^M$ elements, and give a method to build such an ensemble given a strategy. Let  $s$ be a strategy and $U_M$ such an ensemble that contains $s$. Then let $\overline{U}_M$ be such that for all $t$ in $U_M$, $\overline{t}$ belongs to $\overline{U}_M$. All elements of $U_M$, except $s$ itself are uncorrelated with $s$, as do all elements of $\overline{U}_M$, except $\overline{s}$. So, given a strategy $s$, there are $2\cdot 2^M-2$ strategies uncorrelated with $s$, and, of course, one totally anti-correlated. This property clarifies the number of strategies which are really different : it is the cardinal of $V_M=U_M\cup\overline{U}_M$, $2\cdot 2^M$. The ensemble $V_M$ can be called the reduced strategy space. Note that given a strategy $s$ there are $2^{2^M-1}$ $U_M$ that contain $s$, but that $V_M$ is unique in the sense that if $t\in V_M$ and $t\in {V'}_M$, then $V_M={V'}_M$. Of course, a given strategy always belongs to such an ensemble, related to $V_M$ by a given number of reflections. In other words $V_M$ defines a geometrical structure on the hypercube $H_M$.

It is clear now why the quantity $\frac{N}{2^M}$ is a fundamental parameter : it is proportional to the number of drawn strategies over the cardinal of $V_M$, i.e. to the inverse of the density of the system in the reduced strategy space. Therefore, the right parameter should be $\frac{2\cdot 2^M}{S N}$.

\begin{center}
	\begin{figure}
		\label{s2nUM}
		\psfig{file=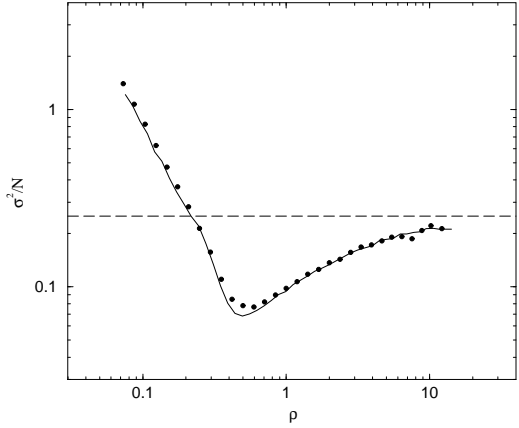,width=8cm}
		\caption{$\sigma^2/N$ versus $\rho$ obtained by drawing all the strategies from $V_M$ ($M=8$, $S=2$). The continuous line shows the original $\sigma^2/N$, and the dashed line indicates the performance of the random case. The two curves are very close together, testifying the fundamental role of $V_M$} 
	\end{figure}
\end{center}

The fundamental role of $V_M$ is shown by the following experience. Since it is easy to construct $V_M$, we can model our model by forcing the players to draw their strategies from a given $V_M$. We plot $\sigma^2/N$ versus $\rho$ (see figure \ref{s2nUM}). One see that $\sigma^2/N$ is really very close to the original one. What do we learn from this comparison ? 
First in the crowded phase ($\rho\ll\rho_c$), we saw that $\sigma^2/N^2\propto 1/2^M$. If $M$ is constant, a system that only differs from another one by $N$ is just a dilatation of the latter. When all strategies belong to $V_M$,  $\frac{S\ N}{2\cdot 2^{M}}$ is the average number of times each strategy has been drawn, thus the cause of the dilatation is obvious.

 Let us now discuss the better-than-random phase. It is tractable analytically in the $\rho\gg 1$ limit, since two strategies are either the same, or uncorrelated, or anti-correlated, which implies that two players are not bound together unless they have identical or anti-correlated strategies. Suppose that $S=2$; if $N\ll 2^M$, it is easy to compute the average maximal number of anti-correlated strategies, $\moy{l}$. Under the hypothesis that a strategy is drawn at most once and that a player cannot possess two anti-correlated strategies, the joint probability that the system can have $l$ pairs of anti-correlated strategies and that $n$ drawn strategies belong to $U_M$ is given by 
\bea
p(l\cup n)&=&\frac{C^{n}_l\ C^{2^M-n}_{\overline{n}-l}\ C^{2N}_{n}}{C^{2^M}_{\overline{n}}\ 2^{2N}}\ \theta(n-l)\nonumber \\
&=&\frac{C^{n}_l\ C^{2^M-n}_{2N-n-l}\ C^{2N}_{n}}{C^{2^M}_{2N-n}\ 2^{2N}}\ \theta(n-l),
\eea

where $n$, respectively. $\overline{n}$, is the number of strategies belonging to $U_M$, respectively. $\overline{U}_M$ (of course $n+\overline{n}=2 N$), and $\theta(x)$ is the Heavyside function. Thus the average number of anti-correlated strategy in the system is equal to

\bea\label{moyl}
\moy{l}&=&\sum^{N}_{n=0}\frac{C^{2N}_{n}}{C^{2^M}_{2N-n}\ 2^{2N}}\sum^{n}_{l=1}l\ C^{n}_l\ C^{2^M-n}_{2N-n-l}\nonumber\\
	&&+\sum^{2N}_{n=N+1}\frac{C^{2N}_{n}}{C^{2^M}_{2N-n}\ 2^{2N}}\sum^{2N-n}_{l=1}l\ C^{n}_l\ C^{2^M-n}_{2N-n-l} \nonumber \\
	&=&\frac{N(N-1/2)}{2^M}.
\eea

Obtaining $\sigma^2/N$ is now straightforward. If the system consists of $N$ uncorrelated players, $\sigma^2/N=1/4$. Now, if there are $\moy{l}$ pairs of anti-correlated players, only $N-2\moy{l}$ players are still uncorrelated, and 

\be\label{sigma2l}
\frac{\sigma^2}{N}=\frac{1}{4}-\frac{\moy{l}}{2N}. 
\ee

Here,
\be
	\frac{\sigma^2_{th}}{N}=\frac{1}{4}-\frac{N-1/2}{2\cdot 2^M}.
\ee

\begin{center}
	\begin{figure}		
		\psfig{file=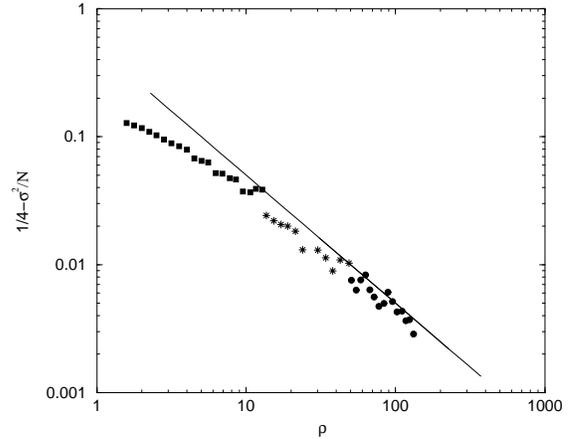,width=8cm}
		\caption{\label{courbeth}Theoretical (continuous line) and original $1/4-\sigma^2/N$ versus $\rho=2^M/N$ for $M=8$ (squares), $M=10$ (stars) and $M=12$ (circles) ($S=2$). The fluctuations are quite important due to the small number of players. One sees the approximation is good if $N\ll 2^M$} 
	\end{figure}
\end{center}

Note that $\sigma^2_{th}/N$ tends to 1/4 like $N/2^M$. Since $\moy{l}$ is the average maximal number of anti-correlated strategies, $\sigma^2_{th}/N$ is a kind of lower bound for $\sigma^2/N$ as it appears on figure \ref{courbeth}, on which one can see that this approximation holds for $N\ll 2^M$. One can also analytically find that $\sigma^2/N$ linearly depends on $\moy{d}$ at least in this region. Indeed, using equation \eqref{sigma2l}, the fact that

\be
\moy{d}=\frac{1}{2}\left(1+\frac{\moy{l}}{N-1}\right),
\ee

  and dropping the terms of order $1/N^2$ yield to the equation
\be\label{sigmadist}
\frac{\sigma^2}{N}\simeq\frac{1}{4}-C\left(\moy{d}-\frac{1}{2} \right),
\ee

where $C$ is a constant equal to 1 in this case, but depends at least on $M$ and $N$ in the original model's case; for instance, $C\simeq0.0302$ on figure \ref{distsigma}. Equation \eqref{sigmadist} indicates that $\sigma^2/N=1/4$ if $\moy{d}=1/2$ and that $\sigma^2/N<1/4$ as soon as $\moy{d}>1/2$ (see figures \ref{deffZ} and \ref{s2Ns}). One might find analytical relationships in the other regions by allowing a strategy to be drawn several times.

\section{Influence of identical players}

\begin{center}
	\begin{figure}		
		\psfig{file=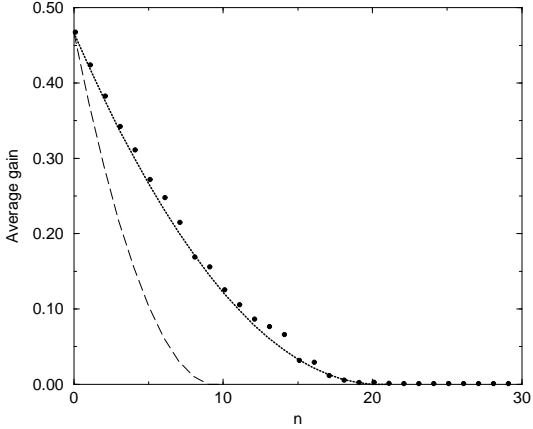,width=8cm}
		\caption{\label{gainblock}Average gain of $n$ identical players against $n/N$ (circles),  $\moy{g}_0-B(n)$ with $w_m=38$ (dashed line) and $w_m=30$ (dotted line) ($N=101$, $M=10$, $S=2$, NIT=5000).} 
	\end{figure}
\end{center}

In this section, we study the effect of cloning a player several times on the system and on the player. One starts with $n$ players having exactly the same strategies and all virtual values set to zero.  Figure \ref{gainblock} shows the average gain of such players against their relative number in the system $n/N$ -- note that one keeps $N$ constant. One sees that their gain is well fitted by a quadratic function and is equal to zero at $n_m=\simeq 20$. 

\begin{center}
	\begin{figure}		
		\psfig{file=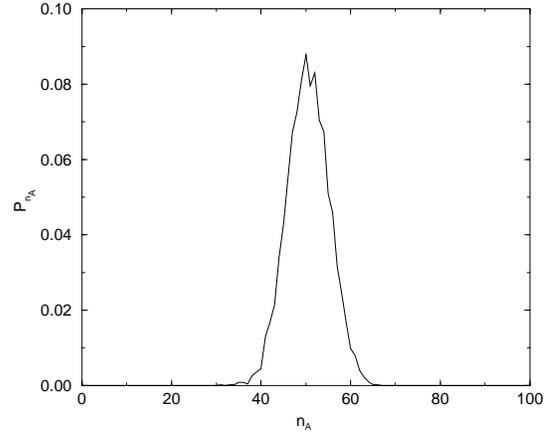,width=8cm}
		\caption{\label{orighisto}Probability distribution of attendance at side $A$, $P_{n_A}$ (same parameters as on figure \ref{gainblock}).} 
	\end{figure}
\end{center}

Calculating the influence of $n$ on the identical players is easy : suppose that at a given time step the cloned player wins if $n=0$; let $w$ be the number of winners ; if $w+n < N/2$, nothing changes for them. But if $w+n> N/2$, the identical players and those who made the same choice lose. That means in particular that the cloned player wins less than before. One can estimate the prejudice $B$ for the cloned player :

\bea
B(n)&=&\sum_{w=0}^{(N-1)/2} P(w)\moy{g}_0\ \theta(w+n-N/2)\nonumber\\
&=&\moy{g}_0\sum_{w=(N+1)/2-n}^{(N-1)/2}P(w),
\eea

where $\moy{g}_0$ is the average gain of the cloned player when $n=0$ and $P(w)$ is the probability that the system gets $w$ winners. Of course one does not know $P(w)$, but figure \ref{orighisto} shows that one can well approximate $P(w)$ by a linear function, 

\be
P(w)\simeq\left\{\begin{array}{ll}	0&\textrm{if $w<w_{m}$}\\
					aw+b&\textrm{if $w_{m}\le w < N/2$}
		\end{array}\right. ,
\ee

where $a$ and $b$ are such that $P(w_{m})=0$ and $P$ is normalized, that is, $b=-a w_{m}$ and 

\be
a=\frac{2}{\left[\displaystyle\frac{N^2-1}{4}-w_{m}(N-w_{m})\right]}\ .
\ee

 Thus, 

\be
B(n)\simeq \frac{a}{2}\left[-n^2+\left(\frac{N}{2}-w_m\right)n\right]\moy{g}_0.
\ee

We eventually define

\be
\moy{g}(n)=\moy{g}_0\left(1-B(n) \right).
\ee

Before one can plot $\moy{g}(n)$, one must find the numerical value of $w_m$. One has two choices. The first one is the mean field way : one suppose that the behavior of the remaining non-cloned players does not differ from the $n=0$ case, and one tries to map the form of $P_{n_A}$ by taking $w_m=38$ ; the resulting $\moy{g}(n)$ is the dashed line on figure \ref{gainblock} and apperas to be too pessimistic. The second one is to consider that the adaptative way by whom all the players, including the identical ones,  use their strategies allows to take $w_m$ such that $P_{n_A}(x\le w_m)=0$, that is, $w_m=30$ ; the result is the dotted line on the same figure and is very close to the experimental data.

\begin{center}
	\begin{figure}		
		\psfig{file=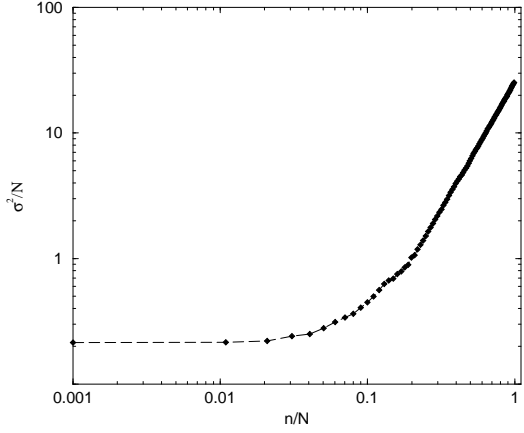,width=8cm}
		\caption{\label{sigmablock}$\sigma^2/N$ against $n/N$. The right part grows like $\left(\frac{n}{N}\right)^2$.} 
	\end{figure}
\end{center}

\begin{center}
	\begin{figure}		
		\psfig{file=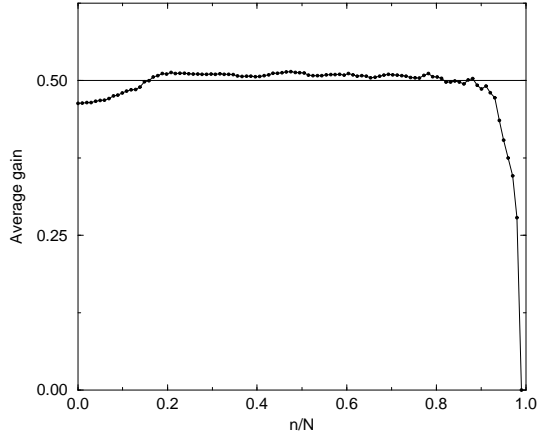,width=8cm}
		\caption{\label{gainpasblock}Average gain of the non-cloned players} 
\end{figure}
\end{center}

If one plots $\sigma^2/N$ against $n/N$ (figure \ref{sigmablock}), one sees that $\sigma^2/N$ remains roughly constant when $n/N$ is small, then grows up. If $n/N>w_m$, $\sigma^2/N$ grows like $(n/N)^2$ ; indeed, since the identical players always lose, the number of losers grows linearly with $n$. One the other hand, figure \ref{gainpasblock} shows the average gain of the non-cloned players. One note that they take advantage of the situation as soon as $n>0$ and that the average gain grows beyond 1/2 and stays roughly constant as soon as the average gain of the identical players fall to zero ; it is a consequence of the adaptative use of the strategies

\begin{center}
	\begin{figure}		
		\psfig{file=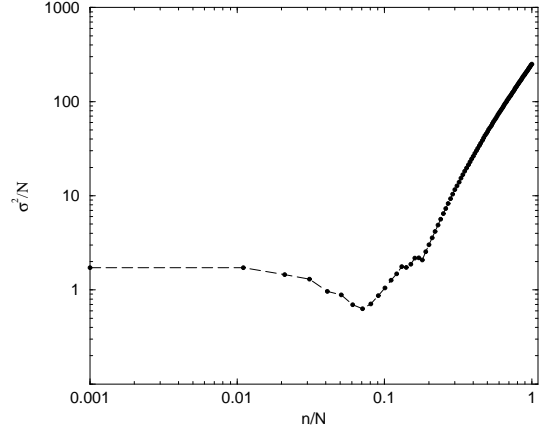,width=8cm}
		\caption{\label{sigmablocktasse}$\sigma^2/N$ against $n/N$. Note that $\sigma^2/N$ has a minimum.} 
	\end{figure}
\end{center}

Note that the studied system was such that $\rho>\rho_c$. If $\rho<\rho_c$, the prejudice $B(n)$ is still a growing with $n$ function, but  approximating $P(w)$ is more difficult since the histogram of the attendance at side $A$ has three peaks, and accordingly $B(n)$ is more complicate. One sees on figure \ref{sigmablocktasse} that $\sigma^2/N$ has minimum, because the actual number of different players is $N-n$, consequently $\rho$ increases when $n$ grows.

\section{Efficiency}

\begin{center}
	\begin{figure}		
		\psfig{file=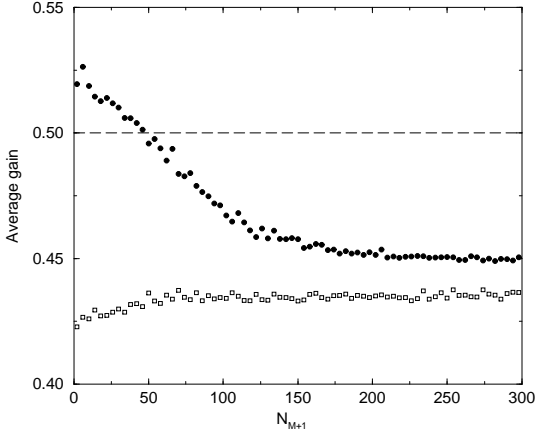,width=8cm}
		\caption{\label{compgain}Average gain of the profiteers (circles) and of the victims (squares) against the number of profiteers ($N_M$=101, $N_{M+1}=2,\ldots,300$, $M=3$, $S=2$).} 
	\end{figure}
\end{center}

\begin{center}
	\begin{figure}		
		\psfig{file=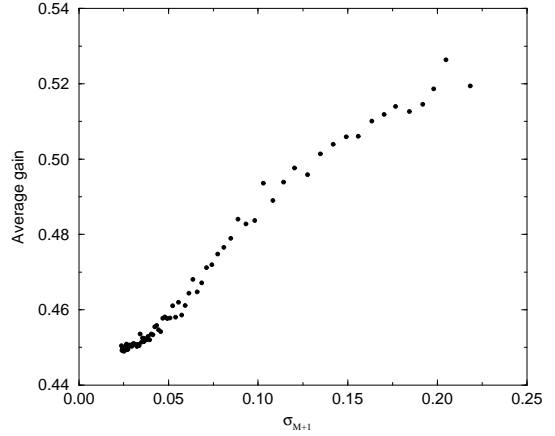,width=8cm}
		\caption{\label{effgain}Monotonical relationship between the average gain of the profiteers and $\sigma_{M+1}$.} 
	\end{figure}
\end{center}

One interesting feature of our model is the following : suppose that $\rho<\rho_c$, then the game is efficient in the sense that given an information,  the two sides win with equal probability at next time step; reversly, if $\rho>\rho_c$, the game is inefficient. Even more, suppose that $N_M$ players have the same memory $M$ and that the game is efficient, then the game is inefficient for hypothetical players who would have a memory equal to $M+1$ \cite{Savit}. Since the latter can take advantage from the situation, it is interesting to add $N_{M+1}$ of them to the game in order to see what they can win and when the inefficieny disappears for the $M+1$ memory. On figure \ref{compgain} one plots the average gain of the profiteers and of the victims. The latter grows a little bit and then stays roughly constant; indeed, the cleverness of the profiteers compensates the bad performance of the victims. The gain of the profiteers starts from over 1/2, linearly decreases until $N_{M+1}\simeq N_M$, falling below 1/2 at $N_{M+1}=N_{M}/2$,  and then decreases less than linearly, but it is not clear whether their gain eventually stabilizes above the victims' one. Therefore, one characterizes the inefficiency of the system for players with memory $M$ by $\sigma_M$ defined by 

\be
\sigma^2_M=\frac{1}{2^{M}}\sum^{2^{M}}_{i=1}(P(A|i)-1/2)^2,
\ee

where $i$ an information of length $M$ and $P(A|i)$ is the conditionnal probability that the winning side will be $A$ at next time step given the information $i$. If $\sigma_M=0$, the system is efficient. When one plots $N_{M+1}\sigma_{M+1}$ against $N_{M+1}$, one sees that $\sigma_{M+1}$ behave like $1/N_{M+1}$ when $N_{M+1}$ is roughly greater than $N_M$, that is, $\sigma_{M+1}$ decreases very slowly. That indicates that the profiteers can always take advantage of the victims. Finally, figure \ref{effgain} shows that the average gain of the profiteers is monotonically related to $\sigma_{M+1}$, since the inefficiency $\sigma_{M+1}$ is a measure of the opportunist gains the profiteers can get.

\begin{center}
	\begin{figure}		
		\psfig{file=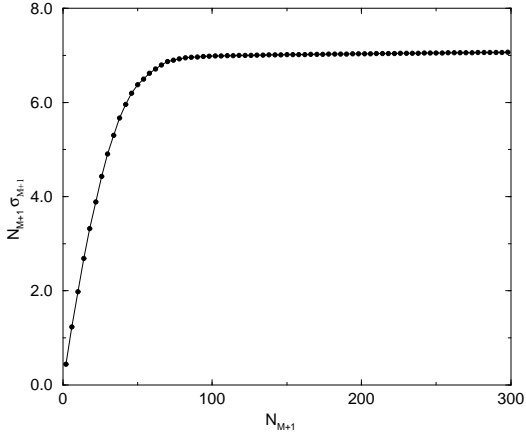,width=8cm}
		\caption{\label{profsigma}$N_{M+1}\sigma_{M+1}$ against $N_{M+1}$, showing that $\sigma_{M+1}\sim 1/N_{M+1}$ when $N_{M+1}>N_{M}$.} 
	\end{figure}
\end{center}

\section{Extensions of the model}
 
Although being simple, our model is very rich and allows a lot of variations. For instance, the {\em minority game} seems to be a special case of the {\em bar problem}, but one should not think the binary nature of our game prevents the system from learning a different from $N/2$ threshold. Indeed, one just has to modify the payoff by breaking the symmetry between the sides $A$ and $B$ : the players at side $A$ win a point if they are less than $\alpha N$ ($0<\alpha<1$); reversely, if there are more than $\alpha N$ players at side $A$, those who choose the side $B$ win. The results are roughly the same as before, except that the game is no more symmetrical. On the other hand, one should argue that our model would be more realistic if the oldest virtual gains of the strategies were not kept in memory. One can modify the model in such a way that only the $T$ newest virtual gains are taken into account to determine which strategy a player uses. This modification does not change the properties a lot, in particular figure \ref{s2Ns}, if $T$ is not too small with respect to $M$, lets the model be Markovian of order $T$ and might help to exactly solve the model. One can also imagine weighing the virtual gains, that is, multiplying the reward gained $t$ time steps ago by $c^t$ where $0<c<1$ in order to include a progressive oblivion; one might find a critical value of $c$, as in the prisoner's dilemma \cite{Axel}.

\section{Conclusion}

We have reviewed some interesting properties of our model; in particular, we have given a geometrical explanation of the phase transition, which has allowed us to find an analytical expression of $\sigma^2/N$ in the $\rho\gg 1$ region. Nevertheless, an analytical solution of the whole model is still missing, being rather hard to find due to the adaptative use of the strategies. Consequently, one could be interested in another approach proposed by one of us \cite{YCZ} : one only gives one binary strategy to each player; there are active and passive players; the active ones play the minority game, the passive ones observe, waiting for the best moment to play. This model might reproduce some of the properties of ours and permit an easier analytical approach. 

This work has been supported in part by the Swiss National Foundation through the Grant No. 20-46918.96.

\appendix
\section{}

In this appendix we show that an ensemble of strategies whose all elements are uncorrelated contains at most $2^M$ elements. The proof also gives a method to build such an ensemble given a strategy $s\in H_M$. Without loss of generality, we consider $s(i)=0$, $i=1,\dots,2^M$. We note $U_{M}$ a maximal subset of $H_M$ such that all its elements are mutually uncorrelated and that it contains $s$; note that $U_M$ is not unique (see above). It is easy to build $U_{M}$ from $U_{{M-1}}$. If $a,\ b\in H_{{M-1}}$ one defines the direct product $a\otimes b$ by 

\bdm
(a\otimes b)(i)=\left\{ \begin{array}{ll}
				a(i) 		&\textrm{if $1\le i\le 2^{M-1}$}\\
				b(i-2^{M-1}) 	&\textrm{if $2^{M-1}<i\le 2^{M}$}. 
			\end{array}\right.
\edm

This product actually extends $a\in H_{M-1}$ in $H_M$ by appending the components of $b$. The ensemble $U_M$ is simply obtained from $U_{M-1}$ by taking all $a\in U_{M-1}$, and putting $a\otimes a$ and $a \otimes \overline{a}$ in $U_M$. It is easy to see that all elements of $U_M$ are mutually uncorrelated.  Since $U_0=\{(0)\}$, $U_M$ contains $2^M$ elements and one proves that its size is maximal by recurrence :

(i) clearly, $U_1=\{(0,0);(0,1)\}$ is as large as possible.

(ii) suppose that the size of $U_{M-1}$ is maximal.

(iii) let $h\in H_M$; suppose that $\forall\ u\in U_M$ ($u\neq h$), $d_M(h,u)=1/2$. Consider now one $a \in U_M$, $a \neq h$; one can decompose $h$ in $h_1\otimes h_2$ where $h_1,\ h_2 \in H_{M-1}$, and $a$ in $b\otimes c$ where $b,c\in U_{M-1}$; remember that $c=b$ or $c=\overline{b}$. Since $b\otimes c$ and $b\otimes \overline{c}\in U_M$, 
\bea
d_M(h,a)&=&\frac{1}{2}\pr{d_{M-1}(h_1,b)+d_{M-1}(h_2,c)}\nonumber \\
	&=&\frac{1}{2}\pr{d_{M-1}(h_1,b)+d_{M-1}(h_2,\overline{c})}\nonumber \\
	&=&\frac{1}{2}\pr{d_{M-1}(h_1,b)+1-d_{M-1}(h_2,c)}=1/2
\eea

Thus $d_{M-1}(h_2,c)=1/2$, then $h_2\in U_{M-1}$, as does $h_1$, that is, $h\in U_M$. In other words, $U_M$ contains at most $2^M$ elements.

\bibliography{minorite}
\end{document}